\def\year{2021}\relax
\documentclass[letterpaper]{article} 
\usepackage{aaai21}  
\usepackage{times}  
\usepackage{helvet} 
\usepackage{courier}  
\usepackage[hyphens]{url}  
\usepackage{graphicx} 
\usepackage{natbib}  
\usepackage{caption} 
\usepackage{amsmath}
\usepackage{amssymb}
\usepackage{booktabs}
\usepackage{algorithm,algorithmic}
\usepackage{amsmath,pifont}
\usepackage{bm}
\usepackage{adjustbox}
\newcommand{\cmark}{\ding{51}}
\newcommand{\xmark}{\ding{55}}

\usepackage{xcolor}

\usepackage{hyperref}
\hypersetup{
    colorlinks=true,
    citecolor=black,
    urlcolor=blue,
}

\title{Context Matters: Graph-based Self-supervised Representation Learning for Medical Images}

\author{

    Li Sun\footnote{Equal contribution}, Ke Yu\footnotemark[\value{footnote}], and Kayhan Batmanghelich
}
\affiliations{

    University of Pittsburgh, USA\\
  \{lis118, key44, kayhan\}@pitt.edu
}

\begin{document}

\maketitle

\begin{abstract}

Supervised learning method requires a large volume of annotated datasets. Collecting such datasets is time-consuming and expensive. Until now, very few annotated COVID-19 imaging datasets are available. Although self-supervised learning enables us to bootstrap the training by exploiting  unlabeled data, the generic self-supervised methods for natural images do not sufficiently incorporate the \emph{context}. For medical images, a desirable method should be sensitive enough to detect deviation from normal-appearing tissue of each anatomical region; here, anatomy is the context. We introduce a novel approach with two levels of self-supervised representation learning objectives: one on the regional anatomical level and another on the patient-level. We use graph neural networks to incorporate the relationship between different anatomical regions. The structure of the graph is informed by anatomical correspondences between each patient and an anatomical atlas. In addition, the graph representation has the advantage of handling any arbitrarily sized image in full resolution. Experiments on large-scale Computer Tomography (CT) datasets of lung images show that our approach compares favorably to baseline methods that do not account for the context. We use the learnt embedding to quantify the clinical progression of COVID-19 and show that our method generalizes well to COVID-19 patients from different hospitals. Qualitative results suggest that our model can identify clinically relevant regions in the images.
\end{abstract}

\section{Introduction}
\label{sec:introduction}
While deep neural network trained by the supervised approach has made breakthroughs in many areas, its performance relies heavily on large-scale annotated datasets. Learning informative representation without human-crafted labels has achieved great success in the computer vision domain~\cite{wu2018unsupervised,chen2020simple,he2020momentum}. Importantly, the unsupervised approach has the capability of learning robust representation since the features are not optimized towards solving a single supervised task. Self-supervised learning has emerged as a powerful way of unsupervised learning. It derives input and label from an unlabeled dataset and formulates heuristics-based pretext tasks to train a model. Contrastive learning, a more principled variant of self-supervised learning, relies on instance discrimination~\cite{wu2018unsupervised} or contrastive predictive coding (CPC)~\cite{oord2018representation}. It has achieved state-of-the-art performance in many aspects, and can produce features that are comparable to those produced by supervised methods~\cite{he2020momentum,chen2020simple}. However, for medical images, the generic formulation of self-supervised learning doesn't incorporate domain-specific anatomical context.

For medical imaging analysis, a large-scale annotated dataset is rarely available, especially for emerging diseases, such as COVID-19. However, there are lots of unlabeled data available. Thus, self-supervised pre-training presents an appealing solution in this domain. There are some existing works that focus on self-supervised methods for learning image-level representations. \cite{chen2019self} proposed to learn image semantic features by restoring computerized tomography (CT) images from the corrupted input images. \cite{taleb2019multimodal} introduced puzzle-solving proxy tasks using multi-modal magnetic resonance images (MRI) scans for representation learning. \cite{bai2019self} proposed to learn cardiac MR image features from anatomical positions automatically defined by cardiac chamber view planes.  Despite their success, current methods suffer from two challenges: (1) These methods do not 
account for anatomical context. For example, the learned representation is invariant with respect to body landmarks which are highly informative for clinicians.
(2) Current methods rely on fix-sized input. The dimensions of raw volumetric medical images can vary across scans due to the differences in subjects' bodies, machine types, and operation protocols. The typical approach for pre-processing natural images is to either resize the image or crop it to the same dimensions, because the convolutional neural network (CNN) can only handle fixed dimensional input. However, both approaches can be problematic for medical images. Taking chest CT for example, reshaping voxels in a CT image may cause distortion to the lung~\cite{singla2018subject2vec}, and cropping images may introduce undesired artifacts, such as discounting the lung volume.

To address the challenges discussed above, we propose a novel method for context-aware unsupervised representation learning on volumetric medical images. First, in order to incorporate context information, we represent a 3D image as a graph of patches centered at landmarks defined by an anatomical atlas. The graph structure is informed by anatomical correspondences between the subject's image and the atlas image using registration.
Second, to handle different sized images, we propose a hierarchical model which learns anatomy-specific representations at the patch level and learns subject-specific representations at the graph level. 
On the patch level, we use a conditional encoder to integrate the local region's texture and the anatomical location.
On the graph level, we use a graph convolutional network (GCN) to incorporate the relationship between different anatomical regions. 

Experiments on a publicly available large-scale lung CT dataset of Chronic Obstructive Pulmonary Disease (COPD) show that our method compares favorably to other unsupervised baselines and outperforms supervised methods on some metrics. We also show that features learned by our proposed method outperform other baselines in staging lung tissue abnormalities related to COVID-19. Our results show that the pre-trained features on large-scale lung CT datasets are generalizable and transfer well to COVID-19 patients from different hospitals. Our code and supplementary material are available at https://github.com/batmanlab/Context\_Aware\_SSL

In summary, we make the following contributions:
\begin{itemize}
\item We introduce a context-aware self-supervised representation learning method for volumetric medical images. The context is provided by both local anatomical profiles and graph-based relationship.
\item We introduce a hierarchical model that can learn both local textural features on patch and global contextual features on graph. The multi-scale approach enables us to handle arbitrary sized images in full resolution. 
\item We demonstrate that features extracted from lung CT scans with our method have a superior performance in staging lung tissue abnormalities related with COVID-19 and transfer well to COVID-19 patients from different hospitals.
\item We propose a method that provides task-specific explanation for the predicted outcome. The heatmap results suggest that our model can identify clinically relevant regions in the images.
\end{itemize}


\section{Method}
\label{sec:method}
\begin{figure*}[htp]
\centering
    \includegraphics[width = \textwidth]
    {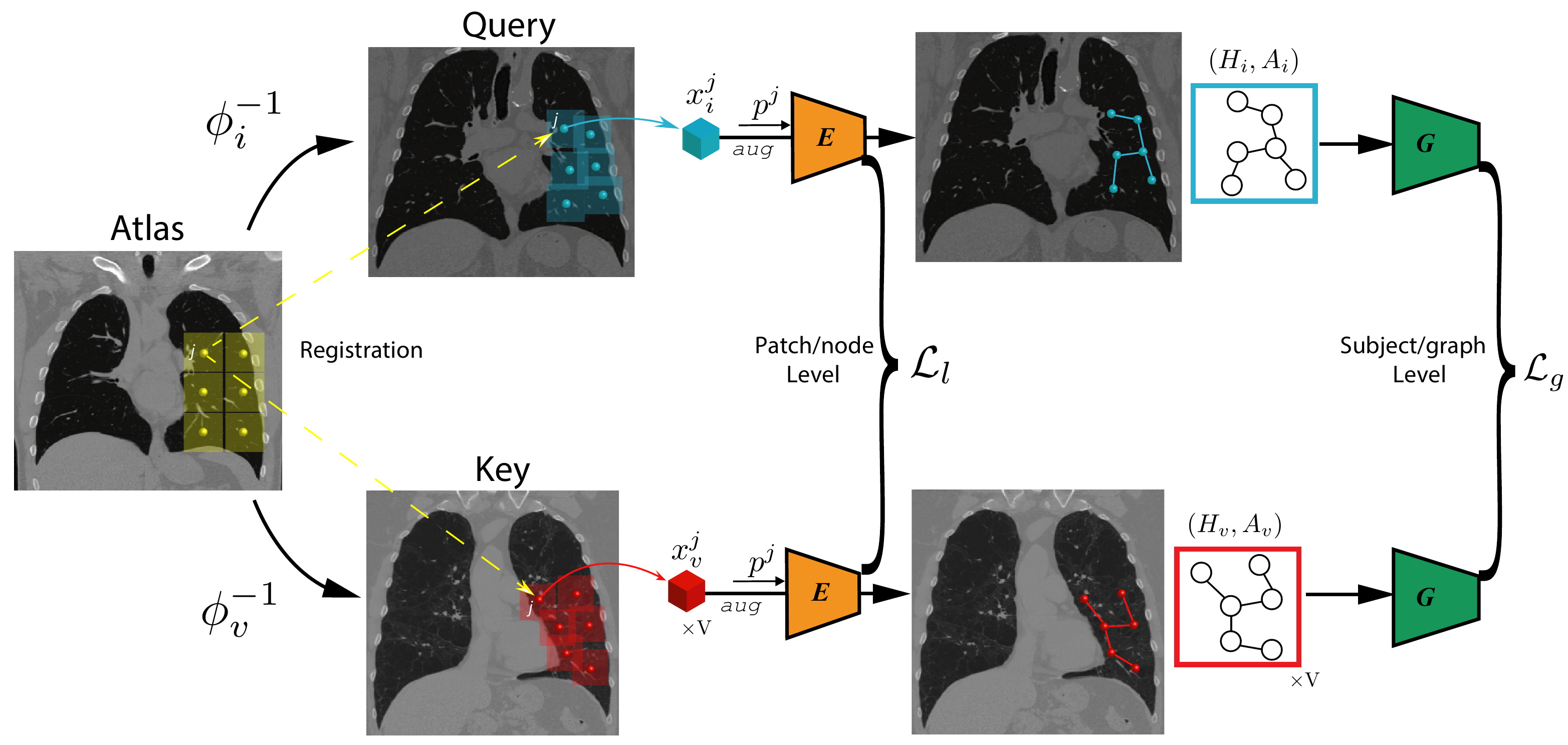}
    \caption{Schematic diagram of the proposed method. We represent every image as a graph of patches. The context is 
    imposed by anatomical correspondences among patients via registration and graph-based hierarchical model used to incorporate the relationship between different anatomical regions. We use a conditional encoder $E(\cdot,\cdot)$ to learn patch-level textural features and use graph convolutional network $G(\cdot,\cdot)$ to learn graph-level representation through contrastive learning objectives. 
    The detailed architecture of the networks are presented in \textbf{Supplementary Material}. 
    }
    \label{fig:main}
\end{figure*}

Our method views images of every patient as a set of nodes where nodes correspond to image patches covering the lung region of a patient. Larger lung (image) results in more spread out patches. We use image registration to an anatomical atlas to maintain the anatomical correspondences between nodes. The edge connecting nodes denote neighboring patches after applying the image deformation derived from image registration. Our framework consists of two levels of self-supervised learning, one on the node level (i.e., patch level) and the second one on the graph level (i.e., subject level). In the following, we explain each component separately. The schematic is shown in Fig.~\ref{fig:main}.

\subsection{Constructing Anatomy-aware Graph of Patients}
We use $X_i$ to denote the image of patient $i$.
To define a standard set of anatomical regions, we divide the atlas image into a set of $N$ equally spaced 3D patches with some overlap. We use $\{ p^j \}_j^N$ to denote the center coordinates of the patches in the \emph{Atlas coordinate system}.
We need to map $\{ p^j \}_j^N$ to their corresponding location for each patient. This operation requires \emph{transformations} that densely map every coordinate of the \emph{Atlas} to the coordinate on patients.
To find the transformation, we register a patient's image to an anatomical atlas by solving the following optimization problem:
\begin{equation}
\min_{\phi_i} \hspace{0.15em} Sim(\phi_i(X_i), X_{\text{Atlas}}) + Reg(\phi_i),
\end{equation}
where $Sim$ is a similarity metric (e.g., $\ell_2$ norm), $\phi_i(\cdot)$ is the fitted subject-specific transformation, $Reg(\phi_i)$ is a regularization term to ensure the transformation is smooth enough. The $\phi_i$ maps the coordinate of the patient $i$ to the Atlas. After solving this optimization for each patient, we can use the inverse of this transformation to map $\{ p^j \}_j^N$ to each subject (i.e., $\{ \phi_i^{-1} (p^j) \}$). We use well-established image registration software ANTs~\cite{tustison2014large} to ensure the inverse transformation exists. To avoid clutter in notation, we use $p_i^j$ as a shorthand for $\phi_i^{-1} (p^j)$. In this way, patches with the same index across all subjects map to the same anatomical region on the atlas image:
\begin{equation}
\phi_1(p_1^j) = \phi_2(p_2^j) = \ldots = \phi_i(p_i^j)  = p^j
\end{equation}

To incorporate the relationship between different anatomical regions, we represent an image as a graph of patches (nodes), whose edge connectivity is determined by the Euclidean distance between patches' centers. With a minor abuse of notation, we let $\mathcal{V}_i=\{ x_i^j \}_j^N$ denote the set of patches that cover the lung region of subject $i$.  More formally, the image $X_i$ is represented as $\mathcal{G}_i=(\mathcal{V}_i,\mathcal{E}_i)$, where $\mathcal{V}_i$ is node (patch) information and $\mathcal{E}_i$ denotes the set of edges. We use an adjacency matrix $A_i\in N \times N$ to represent $\mathcal{E}_i$, defined as:
  \begin{equation}
    A_i^{jk}=
    \begin{cases}
      1, & \text{if}\ \text{dist}(  p_i^j,  p_i^k) < \rho \\
      0, & \text{otherwise}
    \end{cases},
  \end{equation}
where $\text{dist}(\cdot,\cdot)$ denotes the Euclidean distance; $p_i^j$ and $ p_i^k\in \mathcal{R}^3$ are the coordinates of centers in patches $x_i^j$ and $x_i^k$, respectively; $\rho$ is the threshold hyper-parameter that controls the density of graph.

\subsection{Learning Patch-level Representation }
Local anatomical variations provide valuable information about the health status of the tissue. For a given anatomical region, a desirable method should be sensitive enough to  detect  deviation  from normal-appearing  tissue. In addition, the anatomical location of lesion plays a role in patients' survival outcomes, and the 
 types of lesion
vary across different anatomical locations in lung. In order to extract anatomy-specific features, we adopt a conditional encoder  $E(\cdot,\cdot)$ that takes both patch $x_i^j$ and its location index $j$
as input. It is composed with a CNN feature extractor $C(\cdot)$ and a MLP head $f_l(\cdot)$, thus we have the encoded patch-level feature:
\begin{equation}
h_i^j=E(x_i^j, j)=f_l(C(x_i^j)\mathbin\Vert p^j),
\end{equation}
where $\mathbin\Vert$ denotes concatenation, .

We adopt the InfoNCE loss~\cite{oord2018representation}, a form of contrastive loss to train the conditional encoder on the patch level:
\begin{equation}
\mathcal{L}_l=-\log\frac{\exp(q_i^j\cdot k_{+}/\tau)}{\exp(q_i^j\cdot k_{+}/\tau)+\sum_{k_{-}} \exp(q_i^j\cdot k_{-}/\tau)},
\end{equation}
where $q_i^j$ denotes the representation of query patch $x_i^j$, $k_{+}$ and $k_{-}$ denotes the representation of the positive and the negative key respectively, and $\tau$ denotes the temperature hyper-parameter. We obtain a positive sample pair by generating two randomly augmented views 
from the same query patch $x_i^j$,
and obtain a negative sample by augmenting the patch $x_v^j$ at the same anatomical region $j$ from a random subject $v$,
specifically:
\begin{align*}
q_i^j&=f_l(C(aug(x_i^j))\mathbin\Vert p^j),
\\
k_+&=f_l(C(aug(x_i^j))\mathbin\Vert p^j),
\\
k_-&=f_l(C(aug(x_v^j))\mathbin\Vert p^j), v\neq i.
\end{align*}

\subsection{Learning Graph-level Representation }
We adopt the Graph Convolutional Network (GCN)~\cite{duvenaud2015convolutional} to summarize the patch-level (anatomy-specific) representation into the graph-level (subject-specific) representation. We consider each patch as one node in the graph, and the subject-specific adjacent matrix determines the connection between nodes.
Specifically, the GCN model $G(\cdot,\cdot)$ takes patch-level representation $H_i$ and adjacency matrix $A_i$ as inputs, and propagates information across the graph to update node-level features:
\begin{equation}
H'_i =\texttt{concat}(\{{h'}_i^j\}_{j=1}^N)= \sigma\left( \hat{D}_i^{-\frac{1}{2}}\hat{A}_i\hat{D}_i^{-\frac{1}{2}}H_iW\right) ,
\end{equation}
where $\hat{A}_i = A_i + I$, $I$ is an identity matrix, $\hat{D}_i$ is a diagonal node degree matrix of $\hat{A}_i$, $H_i=\texttt{concat}(\{h_i^j\}_{j=1}^N)$ is a $N\times F$ matrix containing $F$ features for all $N$ nodes in the image of the subject $i$, and $W$ is a learnable projection matrix, $\sigma$ is a nonlinear activation function.

We then obtain subject-level representation by global average pooling all nodes in the graph followed by a MLP head $f_g$:
\begin{equation}
\label{eq:6}
S_i=f_g(\texttt{Pool}(H'_i)).
\end{equation}

We adopt the InfoNCE loss to train the GCN on the graph level:
\begin{equation}
\mathcal{L}_g=-\log\frac{\exp(r_i\cdot t_{+}/\tau)}{\exp(r_i\cdot t_{+}/\tau)+\sum_{t_{-}} \exp(r_i\cdot t_{-}/\tau)} ,
\end{equation}
where $r_i^j$ denotes the representation of the entire image $X_i$, $t_{+}$ and $t_{-}$ denotes the representation of the positive and the negative key respectively, and $\tau$ denotes the temperature hyper-parameter.
To form a positive pair, we take two views of the same image $X_i$ under random augmentation at patch level. We obtain a negative sample by randomly sample a different image $X_v$, specifically:
\begin{align*}
r_i&=G(\texttt{concat}(\{E(aug(x_i^n), p^n)\}_{n=1}^N), A_i),
\\
t_+&=G(\texttt{concat}(\{E(aug(x_i^n), p^n)\}_{n=1}^N), A_i),
\\
t_-&=G(\texttt{concat}(\{E(aug(x_v^n), p^n)\}_{n=1}^N), A_v), v\neq i.
\end{align*}

\subsection{Overall Model}
The model is trained in an end-to-end fashion by integrating the two InfoNCE losses obtained from patch level and graph level.
We define the overall loss function as follows:
\begin{equation}
    \mathcal{L}  =  \mathcal{L}_l(E) +  \mathcal{L}_g(G).
\end{equation}
Since directly backpropagating gradients from $\mathcal{L}_g(G)$ to the
parameters in the conditional encoder $E$ is unfeasible due to the excessive memory footprint accounting for a large number of patches, we propose an interleaving algorithm that alternates the training between patch level and graph level to solve this issue. The algorithmic description of the method is shown below:
\begin{algorithm}
\caption{Interleaving update algorithm}
\begin{algorithmic}
\STATE \textbf{Require:} Conditional encoder $E(\cdot,\cdot)$, GCN $G(\cdot,\cdot)$.\\
\STATE \textbf{Input:} Image patch $x_i^j$, anatomical landmark $p^j$, adjacency matrix $A_i$.
\STATE
\FOR{step $t=1,T_{max}$}
    \FOR{step $t_l=1,T_{l}$}
        \STATE Randomly sample a batch of $B_l$ subjects
        \FOR{$j=1,N$}
             \STATE $h_i^j$ $\leftarrow$ $E(aug(x_i^j),p^j)$
             \STATE Update $E$ by backpropagating $\mathcal{L}_l$
        \ENDFOR
    \ENDFOR
    \FOR{step $t_g=1,T_{g}$}
        \STATE Randomly sample a batch of $B_g$ subjects
        \STATE $S_i$$\leftarrow$$G(\texttt{concat}(\{E(aug(x_i^n), p^n)\}_{n=1}^N), A_i)$
        \STATE Update $G$ by backpropagating $\mathcal{L}_g$
    \ENDFOR
\ENDFOR
\end{algorithmic}
\end{algorithm}

\subsection{Model Explanation}
Understanding how the model makes predictions is important to build trust in medical imaging analysis. In this section, we propose a method that provides task-specific explanation for the predicted outcome.
Our method is expanded from the class activation maps (CAM) proposed by~\cite{zhou2016learning}. 
Without loss of generality, we assume a Logistic regression model is fitted for a downstream binary classification task (e.g., the presence or absence of a disease) on the extracted subject-level features $S'_i$. The log-odds of the target variable that $Y_i=1$ is:
%
\begin{equation}
\log \frac{\mathbb{P}(Y_i = 1 | S'_i) }{\mathbb{P}(Y_i = 0 | S'_i) } = \beta+WS'_i=\beta+ \frac{1}{N}\sum_{j=1}^{N}\underbrace{  W h_{i}'^j }_{M_i^j} 
\end{equation}
where $S'_i=\texttt{Pool}(H'_i)$, the MLP head $f_g$ in Eq.~\ref{eq:6} is discarded when extracting features for downstream tasks following the practice in~\cite{chen2020simple,chen2020improved}, $\beta$ and $W$ are the learned logistic regression weights. Then we have $M_i^j=W h_{i}'^j$ as the activation score of the anatomical region $j$ to the target classification. We use a sigmoid function to normalize $\{M_i^j\}_j^N$, and use a heatmap to show the discriminative anatomical regions in the image of subject $i$.

\subsection{Implementation Details}
We train the proposed model for 30 epochs. We set the learning rate to be $3\times10^{-2}$. We also employed momentum $= 0.9$ and weight decay $= 1\times10^{-4}$ in the Adam optimizer. The patch size is set as $32\times32\times32$. The batch size at patch level and subject level is set as 128 and 16, respectively. We let the representation dimension $F$ be $128$. The lung region is extracted using lungmask~\cite{hofmanninger2020automatic}. Following the practice in MoCo, we maintain a queue of data samples and use a momentum update scheme to increase the number of negative samples in training; as shown in previous work, it can improve performance of downstream task~\cite{he2020momentum}. The number of negative samples during training is set as $4096$. The data augmentation includes random elastic transform, adding random Gaussian noise, and random contrast adjustment. The temperature $\tau$ is chosen to be $0.2$. There are $581$ patches per subject/graph, this number is determined by both the atlas image size and two hyperparameters, patch size and step size. The experiments are performed on 2 GPUs, each with 16GB memory. The code is available at \href{https://github.com/batmanlab/Context_Aware_SSL}{https://github.com/batmanlab/Context\_Aware\_SSL}.

\section{Related works}
\subsection{Unsupervised Learning}
Unsupervised learning aims to learn meaningful representations without human-annotated data. Most unsupervised learning methods can be classified into generative and discriminative approaches. Generative approaches learn the distribution of data and latent representation by generation. These methods include adversarial learning and autoencoder based methods. However, generating data at pixel space can be computationally intensive, and generating fine detail may not be necessary for learning effective representation. 

Discriminative approaches use pre-text tasks for representation learning. Different from supervised approaches, both the inputs and labels are derived from an unlabeled dataset. Discriminative approaches can be grouped into (1) pre-text tasks based on heuristics, including solving jigsaw puzzles~\cite{noroozi2016unsupervised}, context predication~\cite{doersch2015unsupervised}, colorization~\cite{zhang2016colorful} and (2) contrastive methods. Among them, contrastive methods achieve state-of-the-art performance in many tasks. The core idea of contrastive learning is to bring different views of the same image (called 'positive pairs') closer, and spread representations of views from different images (called 'negative pairs'). The similarity is measured by the dot product in feature space~\cite{wu2018unsupervised}. Previous works have suggested that the performance of contrastive learning relies on large batch size~\cite{chen2020simple} and large number of negative samples~\cite{he2020momentum}. 

\subsection{Representation Learning for Graph}
Graphs are a powerful way of representing entities with arbitrary relational structure~\cite{battaglia2018relational}.
Several algorithms proposed to use random walk-based methods for unsupervised representation learning on the graph~\cite{grover2016node2vec,perozzi2014deepwalk,hamilton2017inductive}. These methods are powerful but rely more on local neighbors than structural information~\cite{ribeiro2017struc2vec}. Graph convolutional network (GCN)~\cite{duvenaud2015convolutional,kipf2016semi} was proposed to generalize convolutional neural networks to work on the graphs.
Recently, Deep Graph Infomax~\cite{velickovic2019deep} was proposed to learn node-level representation by maximizing mutual information between patch representations and corresponding high-level summaries of graphs.

\section{Experiments}
We evaluate the performance of the proposed model on two large-scale datasets of 3D medical images. We compare our model with various baseline methods, including both supervised approaches and unsupervised approaches.

\subsection{Datasets}

The experiments are conducted on three volumetric medical imaging datasets, including the COPDGene dataset~\cite{regan2011genetic}, the MosMed dataset~\cite{morozov2020mosmeddata} and the COVID-19 CT dataset. All images are re-sampled to isotropic $1 mm^3$ resolution. The Hounsfield Units (HU) are mapped to the intensity window of $[-1024,240]$ and then normalized to $[-1,1]$. 

\subsubsection{COPDGene Dataset}
COPD is a lung disease that makes it difficult to breathe. The COPDGene Study~\cite{regan2011genetic} is a multi-center observational study designed to identify the underlying genetic factors of COPD. We use a large set of 3D thorax computerized tomography (CT) images of 9,180 subjects from the COPDGene dataset in our study. 

\subsubsection{MosMed Dataset} We use 3D CT scans of 1,110 subjects from the MosMed dataset~\cite{morozov2020mosmeddata} provided by municipal hospitals in Moscow, Russia. Based on the severity of lung tissues abnormalities related with COVID-19, the images are classified into five severity categories associated with different triage decisions. For example, the patients in the mild category are followed up at home with telemedicine monitoring, while the patients in the critical category are immediately transferred to the intensive care unit.

\subsubsection{COVID-19 CT Dataset}
To verify whether the learned representation can be transferred to COVID-19 patients from other sites, we collect a multi-hospital 3D thorax CT images of COVID-19. The combined dataset has 80 subjects, in which 35 positive subjects are from multiple publicly available COVID-19 datasets~\cite{COVID-19-SegBenchmark,segcovid,zhou2020rapid}, and 45 healthy subjects randomly sampled from the LIDC-IDRI dataset~\cite{armato2011lung} as negative samples.

\subsection{Quantitative Evaluation}
We evaluate the performance of proposed method by using extracted representations of subjects to predict clinically relevant variables. 

\subsubsection{COPDGene dataset}
We first perform self-supervised pre-training with our method on the COPDGene dataset. Then we freeze the extracted subject-level features and use them to train a linear regression model to predict two continuous clinical variables, percent predicted values of Forced Expiratory Volume in one second (\texttt{FEV1pp}) and its ratio with Forced vital capacity (FVC) (\texttt{$\text{FEV}_1 / \text{FVC}$}),  on the the log scale. We report average $R^2$ scores with standard deviations in five-fold cross-validation.
We also train a logistic regression model for each of the six categorical variables, including (1) Global Initiative for Chronic Obstructive Lung Disease (GOLD) score, which is a four-grade categorical value indicating the severity of airflow limitation, (2) Centrilobular emphysema (CLE) visual score, which is a six-grade categorical value indicating the severity of emphysema in centrilobular, (3) Paraseptal emphysema (Para-septal) visual score, which is a three-grade categorical value indicating the severity of paraseptal emphysema, (4) Acute Exacerbation history (AE history), which is a binary variable indicating whether the subject has experienced at least one exacerbation before enrolling in the study, (5) Future Acute Exacerbation (Future AE), which is a binary variable indicating whether the subject has reported experiencing at least one exacerbation at the 5-year longitudinal follow up, (6) Medical Research Council Dyspnea Score (mMRC), which is a five-grade categorical value indicating dyspnea symptom.

We compare the performance of our method against: (1) supervised approaches, including  Subject2Vec~\cite{singla2018subject2vec},  Slice-based CNN~\cite{gonzalez2018disease} and (2) unsupervised approaches, including Models Genesis~\cite{zhou2019models}, MedicalNet~\cite{chen2019med3d}, MoCo (3D implementation)~\cite{he2020momentum}, Divergence-based feature extractor~\cite{schabdach2017likelihood}, K-means algorithm applied to image features extracted from local lung regions~\cite{schabdach2017likelihood}, and Low Attenuation Area (LAA), which is a clinical descriptor.
The evaluation results are shown in Table~\ref{tbl:COPD}. For all results, we report average test accuracy in five-fold cross-validation.
\begin{table*}[t]
 \caption{ Evaluation on COPD dataset}
 \begin{adjustbox}{max width=\textwidth}
 \centering
 \label{tbl:COPD}
  \begin{tabular}{lc|cc|cccccc}
  \toprule
   Method&Supervised&$\log$\texttt{FEV1pp}&$\log$\texttt{$\text{FEV}_1 / \text{FVC}$}&GOLD&CLE&Para-septal&AE History&Future AE&mMRC\\
   \toprule
   Metric&  &\multicolumn{2}{|c|}{R-Square}& \multicolumn{6}{c}{\% Accuracy}  \\
   \midrule
 LAA-950 &\xmark&$0.44_{\pm.02}$&$0.60_{\pm.01}$&$55.8$&$32.9$&$33.3$&$73.8$&$73.8$&$41.6$\\
 K-Means &\xmark&$0.55_{\pm.03}$&$0.68_{\pm.02}$&$57.3$&-&-&-&-&-\\
 Divergence-based &\xmark& $0.58_{\pm.03}$&$0.70_{\pm.02}$&$58.9$&-&-&-&-&-\\
 MedicalNet &\xmark&$0.47_{\pm.10}$&$0.59_{\pm.06}$&$57.0_{\pm1.3}$&$40.3_{\pm1.9}$&$53.1_{\pm0.7}$&$78.7_{\pm1.3}$&$81.4_{\pm1.7}$&$47.9_{\pm1.2}$\\
 ModelsGenesis&\xmark&$0.58_{\pm.01}$&$0.64_{\pm.01}$&$59.5_{\pm2.3}$&$41.8_{\pm1.4}$&$52.7_{\pm0.5}$&$77.8_{\pm0.8}$&$76.7_{\pm1.2}$&$46.0_{\pm1.2}$\\
 MoCo &\xmark& $0.40_{\pm.02}$&$0.49_{\pm.02}$&$52.7_{\pm1.1}$&$36.5_{\pm0.7}$&$52.5_{\pm1.4}$&$78.6_{\pm0.9}$&\bm{$82.0_{\pm1.2}$}&$46.4_{\pm1.7}$\\
 \midrule
2D CNN &\cmark&$0.53$&-&$51.1$&-&-&$60.4$&-&-\\
Subject2Vec &\cmark&\bm{$0.67_{\pm.03}$}&\bm{$0.74_{\pm.01}$}&\bm{$65.4$}&$40.6$&$52.8$&$76.9$&$68.3$&$43.6$\\
 \midrule
 Ours w/o CE&\xmark&$0.56_{\pm.03}$&$0.65_{\pm.03}$&$61.6_{\pm1.2}$&$48.1_{\pm0.4}$&$55.5_{\pm0.8}$&\bm{$78.8_{\pm1.2}$}&$80.8_{\pm1.7}$&$50.4_{\pm1.0}$\\
 Ours w/o Graph&\xmark&$0.60_{\pm.01}$&$0.69_{\pm.01}$&$62.5_{\pm1.0}$&$49.2_{\pm1.1}$&$55.8_{\pm1.3}$&$78.7_{\pm1.5}$&$80.7_{\pm1.7}$&$50.6_{\pm0.8}$\\
 Ours&\xmark&$0.62_{\pm.01}$&$0.70_{\pm.01}$&$63.2_{\pm1.1}$&\bm{$50.4_{\pm1.3}$}&\bm{$56.2_{\pm1.1}$}&\bm{$78.8_{\pm1.3}$}&$81.1_{\pm1.6}$&\bm{$51.0_{\pm1.0}$}\\
  \bottomrule
   \multicolumn{10}{p{.5\textwidth}}{- indicates not reported.}
  \end{tabular}
   \end{adjustbox}
 \end{table*}
 
 The results show that our proposed model outperforms unsupervised baselines
 in all metrics except for Future AE. While MoCo is also a contrastive learning based method, we believe that our proposed method achieves better performance for three reasons: (1) Our method incorporates anatomical context. (2) Since MoCo can only accept fixed-size input, we resize all volumetric images into $256\times256\times256$. In this way, lung shapes may be distorted in the CT images, and fine-details are lost due to down-sampling. In comparison, our model supports images with arbitrary sizes in full resolution by design. (3) Since training CNN model with volumetric images is extremely memory-intensive, we can only train the MoCo model with limited batch size. The small batch size may lead to unstable gradients. In comparison, the interleaving training scheme reduces the usage of memory footprint, thus it allows us to train our model with a much larger batch size.
 
Our method also outperforms supervised methods, including Subject2Vec and 2D CNN, in terms of CLE, Para-septal, AE History, Future AE and mMRC; for the rest clinical variables,
the performance gap of our method is smaller than other unsupervised methods. We believe that the improvement is mainly from the richer context information incorporated by our method.
Subject2Vec uses an unordered set-based representation which does not account for spatial locations of the patches. 2D CNN only uses 2D slices which does not leverage 3D structure. Overall, the results suggest that representation extracted by our model preserves richer information about the disease severity than baselines.

\textbf{Ablation study:} We perform two ablation studies to validate the importance of context provided by anatomy and the relational structure of anatomical regions:
(1) Removing conditional encoding (CE). In this setting, we replace the proposed conditional encoder 
with a conventional encoder which only takes images as input. (2) Removing graph. In this setting, we remove GCN in the model and obtain subject-level representation by  average pooling of all patch/node level representations without propagating information between nodes. As shown in Table~\ref{tbl:COPD}, both types of context contribute significantly to the performance of the final model.
 
 \subsubsection{MosMed dataset}
 We first perform self-supervised pre-training with our method on the MosMed dataset. Then we freeze the extracted patient-level features and train a logistic regression classifier to predict the severity of lung tissue abnormalities related with COVID-19, a five-grade categorical variable based on the on CT findings and other clinical measures. We compare the proposed method with benchmark unsupervised methods, including MedicalNet, ModelsGenesis, MoCo, and one supervised model, 3D CNN model. We use the average test accuracy in five-fold cross-validation as the metric for quantifying prediction performance.
 
 \begin{table}[t]
 \centering
 \caption{ Evaluation on MosMed dataset}
 \label{tbl:RU}
  \begin{tabular}{lcc}
  \toprule
   Method&Supervised&\% Accuracy\\
   \midrule
  MedicalNet &\xmark&$62.1$\\
  ModelsGenesis &\xmark&$62.0$\\
MoCo &\xmark&$62.1$\\
3D CNN &\cmark&$61.2$\\
 \midrule
 Ours w/o CE&\xmark&$63.3$\\
Ours w/o Graph&\xmark&$64.3$\\
Ours&\xmark&\bm{$65.3$}\\
  \bottomrule
  \end{tabular}
 \end{table}
 
 

Table~\ref{tbl:RU} shows that our proposed model outperforms both the unsupervised and supervised baselines. The supervised 3D CNN model performed worse than the other unsupervised methods, suggesting that it might not converge well or become overfitted since the size of the training set is limited. The features extracted by the proposed method show superior performance in staging lung tissue abnormalities related with COVID-19 than those extracted by other unsupervised benchmark models. We believe that the graph-based feature extractor provides additional gains by utilizing the full-resolution CT images than CNN-based feature extractor, which may lose information after resizing or downsampling the raw images. The results of ablation studies support that counting local anatomy and relational structure of different anatomical regions is useful for learning more informative representations for COVID-19 patients.
 
\subsubsection{COVID-19 CT Dataset}
Since the size of COVID-19 CT Dataset is very small (only 80 images are available), we don't train the networks from scratch with this dataset. Instead, we use models pre-trained on the COPDGene dataset and the MosMed dataset to extract patient-level features from the images in the COVID-19 CT Dataset, and train a logistic regression model on top of it to classify COVID-19 patients. We compare the features extracted by the proposed method to the baselines including MedicalNet, ModelsGenesis, MoCo (unsupervised), and 3D CNN (supervised). We report the average test accuracy in five-fold cross-validation.

 \begin{table*}[t]
 \caption{ Evaluation on COVID-19 CT dataset}

 \centering
 \label{tbl:COVID}
  \begin{tabular}{lcc}
  \toprule
   Method&Supervised&\% Accuracy\\
   \midrule
   MedicalNet &\xmark&$85.0$\\
ModelsGenesis (Pretrained on COPDGene) &\xmark&$92.5$\\
ModelsGenesis (Pretrained on MosMed) &\xmark&$88.7$\\
MoCo (Pretrained on COPDGene) &\xmark&$75.0$\\
MoCo (Pretrained on MosMed) &\xmark&$86.3$\\
3D CNN &\cmark&$77.5$\\
 \midrule
Ours (Pretrained on COPDGene)&\xmark&$90.0$\\
Ours (Pretrained on MosMed)&\xmark&\bm{$96.3$}\\
  \bottomrule
  \end{tabular}

 \end{table*}
 
 
 
Table~\ref{tbl:COVID} shows that the features extracted by the proposed model pre-trained on the MosMed dataset perform the best for COVID-19 patient classification. They outperform the features extracted by the same model pre-trained on the COPDGene dataset. We hypothesize that this is because the MosMed dataset contains subjects with COVID-19 related pathological tissue, such as ground glass opacities and mixed attenuation. However, the COPDGene dataset also shows great performance for transfer learning with both the ModelsGenesis model and our model, which shed light on the utility of unlabeled data for COVID-19 analysis. 
 
\subsection{Model Visualization}
\begin{figure}[t]
\centering
    \includegraphics[width = .44\textwidth]
    {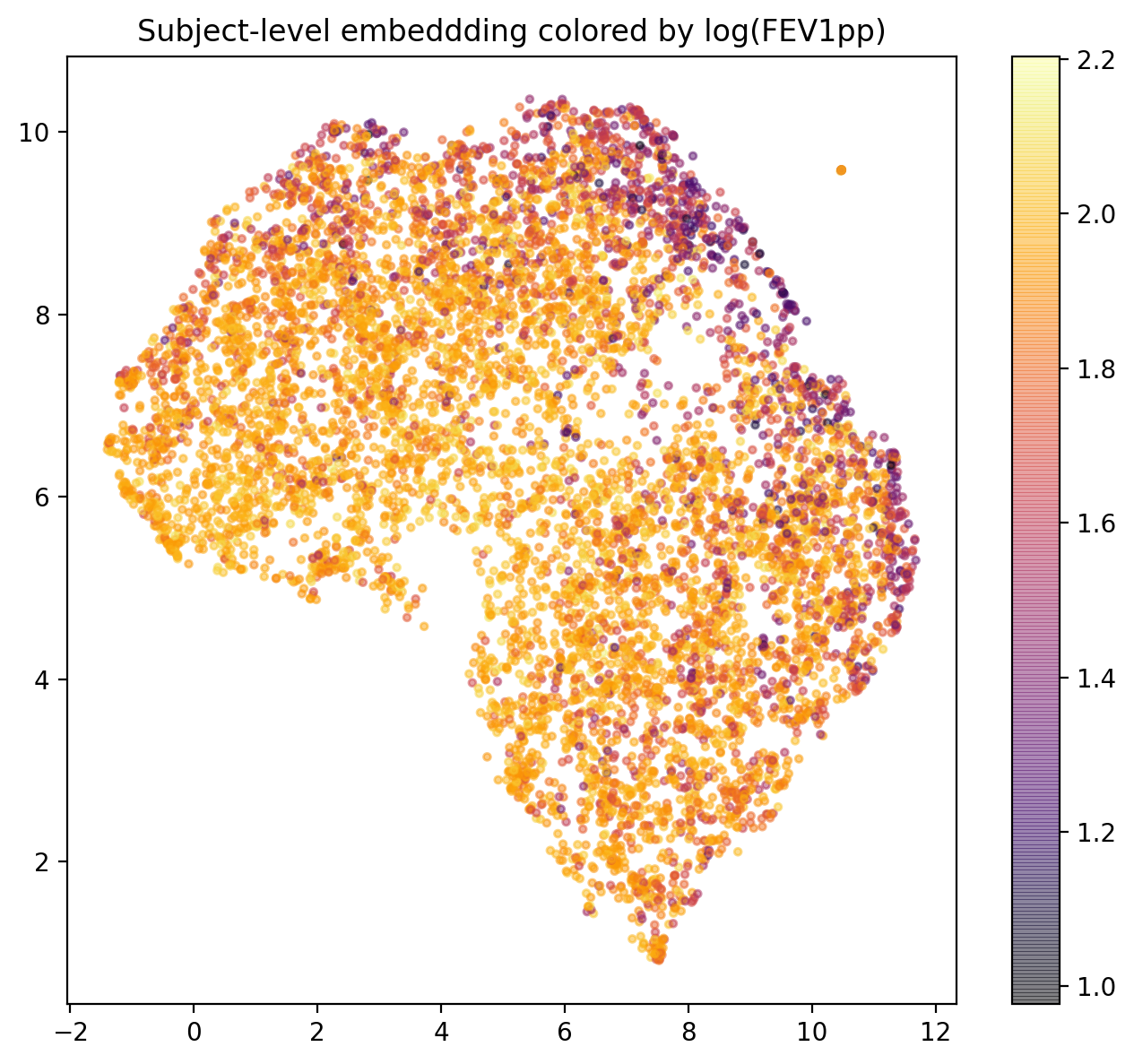}
    \caption{Embedding of subjects in 2D using UMAP. Each dot represents one subject colored by $\log$ \texttt{FEV1pp}. Note that lower $\texttt{FEV1pp}$ value indicates more severe disease.
    }
    \label{fig:fev1}
\end{figure}

To visualize the learned embedding and understand the model's behavior, we use two methods to visualize the model. The first one is embedding visualization,
we use UMAP~\cite{mcinnes2018umap} to visualize the patient-level features extracted on the COPDGene dataset in two dimensions. Figure~\ref{fig:fev1} shows a trend, from lower-left to upper-right, along which the value of \texttt{FEV1pp} decreases or the severity of disease increases.


\begin{figure}[t]
\centering
    \includegraphics[width = .45\textwidth]
    {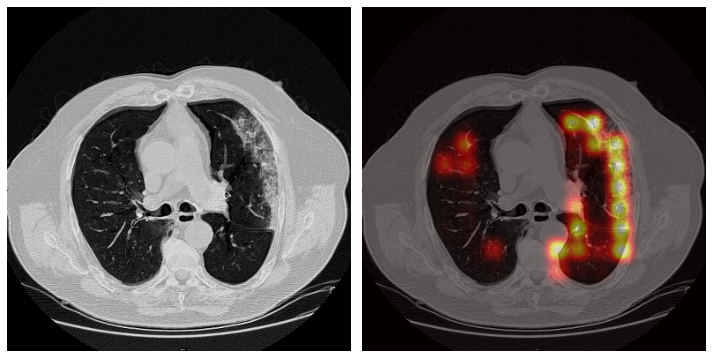}
    \caption{An axial view of the activation heatmap on a COVID-19 positive subject in the COVID-19 CT dataset. Brighter color indicate higher relevance to the disease severity. The figure illustrates that high activation region overlaps with the ground glass opacities.
    }
    \label{fig:covid}
\end{figure}

In addition, we use the model explanation method introduced before to obtain the activation heatmap relevant to the downstream task, COVID-19 classification. Figure~\ref{fig:covid} (left) shows the axial view of the CT image of a COVID-19 positive patient, and Figure~\ref{fig:covid} (right) shows the corresponding activation map. The anatomical regions received high activation scores overlap with the peripheral ground glass opacities on the CT image, which is a known indicator of COVID-19. We also found that activation maps of non-COVID-19 patients usually have no obvious signal, which is expected. This result suggests that our model can highlight the regions that are clinically relevant to the prediction. More examples can be found in \textbf{Supplementary Material}.


\section{Conclusion}
In this paper, we introduce a novel method for context-aware unsupervised representation learning on volumetric medical images. We represent a 3D image as a graph of patches with anatomical correspondences between each patient, and incorporate the relationship between anatomical regions. In addition, we introduced a multi-scale model which includes a conditional encoder for local textural feature extraction and a graph convolutional network for global contextual feature extraction. Moreover, we propose a task-specific method for model explanation. The experiments on multiple datasets demonstrate that our proposed method is effective, generalizable and interpretable.

\bibliography{main.bib}

\end{document}


\maketitle
\section{Network Archtecture}
In the tables below, we show the detailed architectures of conditional encoder $E(\cdot,\cdot)$, including $C(\cdot)$ and $f_l(\cdot)$, and graph convolutional network $G(\cdot,\cdot)$.
\begin{table}[H]
\centering
\caption{Architecture of the $C$ Network}
\begin{adjustbox}{max width=.5\textwidth}
\label{tbl:arch_c}
\begin{tabular}{lcc}

\toprule
Layer
&Filter size, stride
&Output size$(C,D,H,W)$
\\
\midrule
Input
& - & 1$\times$32$\times$32$\times$32
\\
\midrule
Conv3D
& 3$\times$3$\times$3, 1 & 8$\times$32$\times$32$\times$32
\\
BatchNorm+ELU
& - & 8$\times$32$\times$32$\times$32
\\
Conv3D
& 3$\times$3$\times$3, 2 & 8$\times$16$\times$16$\times$16
\\
BatchNorm+ELU
& - & 8$\times$16$\times$16$\times$16
\\
\midrule
Conv3D
& 3$\times$3$\times$3, 1 & 16$\times$16$\times$16$\times$16
\\
BatchNorm+ELU
& - & 16$\times$16$\times$16$\times$16
\\
Conv3D
& 3$\times$3$\times$3, 1 & 16$\times$16$\times$16$\times$16
\\
BatchNorm+ELU
& - & 16$\times$16$\times$16$\times$16
\\
Conv3D
& 3$\times$3$\times$3, 2 & 16$\times$8$\times$8$\times$8
\\
BatchNorm+ELU
& - & 16$\times$8$\times$8$\times$8
\\
\midrule
Conv3D
& 3$\times$3$\times$3, 1 & 32$\times$8$\times$8$\times$8
\\
BatchNorm+ELU
& - & 32$\times$8$\times$8$\times$8
\\
Conv3D
& 3$\times$3$\times$3, 1 & 32$\times$8$\times$8$\times$8
\\
BatchNorm+ELU
& - & 32$\times$8$\times$8$\times$8
\\
Conv3D
& 3$\times$3$\times$3, 2 & 32$\times$4$\times$4$\times$4
\\
BatchNorm+ELU
& - & 32$\times$4$\times$4$\times$4
\\
\midrule
Conv3D
& 3$\times$3$\times$3, 1 & 64$\times$4$\times$4$\times$4
\\
BatchNorm+ELU
& - & 64$\times$4$\times$4$\times$4
\\
Conv3D
& 3$\times$3$\times$3, 1 & 64$\times$4$\times$4$\times$4
\\
BatchNorm+ELU
& - & 64$\times$4$\times$4$\times$4
\\
Conv3D
& 3$\times$3$\times$3, 2 & 64$\times$2$\times$2$\times$2
\\
BatchNorm+ELU
& - & 64$\times$2$\times$2$\times$2
\\
\midrule
Conv3D
& 3$\times$3$\times$3, 1 & 128$\times$2$\times$2$\times$2
\\
BatchNorm+ELU
& - & 128$\times$2$\times$2$\times$2
\\
Conv3D
& 3$\times$3$\times$3, 2 & 128$\times$1$\times$1$\times$1
\\
BatchNorm+ELU
& - & 128$\times$1$\times$1$\times$1
\\
\midrule
Reshape
& - & 1$\times$128
\\
\bottomrule
\end{tabular}
\end{adjustbox}
\end{table}

\begin{table}[H]
\centering
\caption{Architecture of the $f_l$ Network}
\begin{adjustbox}{max width=.5\textwidth}
\label{tbl:arch_fl}
\begin{tabular}{lcc}

\toprule
Layer
&Filter size, stride
&Output size$(C,F)$
\\
\midrule
Input
& - & 1$\times$128,1$\times$3
\\
Concatenation
& - & 1$\times$131
\\
\midrule
Dense
& - & 1$\times$131
\\
ReLU
& - & 1$\times$131
\\
\midrule
Dense
& - & 1$\times$131
\\
ReLU
& - & 1$\times$131
\\
\midrule
Dense
& - & 1$\times$128
\\
\bottomrule
\end{tabular}
\end{adjustbox}
\end{table}

\begin{table}[H]
\centering
\caption{Architecture of the $G$ Network}
\begin{adjustbox}{max width=.5\textwidth}
\label{tbl:arch_g}
\begin{tabular}{lcc}

\toprule
Layer
&Filter size, stride
&Output size$(C,N,F)$
\\
\midrule
Input
& - & 1$\times$581$\times$128
\\
\midrule
GCNLayer
& - & 1$\times$581$\times$128
\\
BatchNorm+ELU
& - & 1$\times$581$\times$128
\\
AveragePooling
& - & 1$\times$1$\times$128
\\
\midrule
Dense
& - & 1$\times$1$\times$128
\\
ReLU
& - & 1$\times$1$\times$128
\\
\midrule
Dense
& - & 1$\times$1$\times$128
\\
ReLU
& - & 1$\times$1$\times$128
\\
\midrule
Dense
& - & 1$\times$1$\times$128
\\
\midrule
Reshape
& - & 1$\times$128
\\
\bottomrule
\end{tabular}
\end{adjustbox}
\end{table}

\section{Implementation Details (cont)}
The patch size is set as $32\times32\times32$. Cosine schedule~\cite{chen2020improved} is used to update the learning rate. For MoCo~\cite{he2020momentum}, we implement a 3D encoder to handle the 3D data and train the model on COPDGene and MosMed dataset.
For ModelsGenesis~\cite{zhou2019models}, we train the model on COPDGene and MosMed dataset with the original setting. For MedicalNet~\cite{chen2019med3d}, since it's training requires segmentation mask, we use pretrained weights provided by the authors.

\section{Model Visualization}
\begin{figure}[t]
\centering
    \includegraphics[width = .49\textwidth]
    {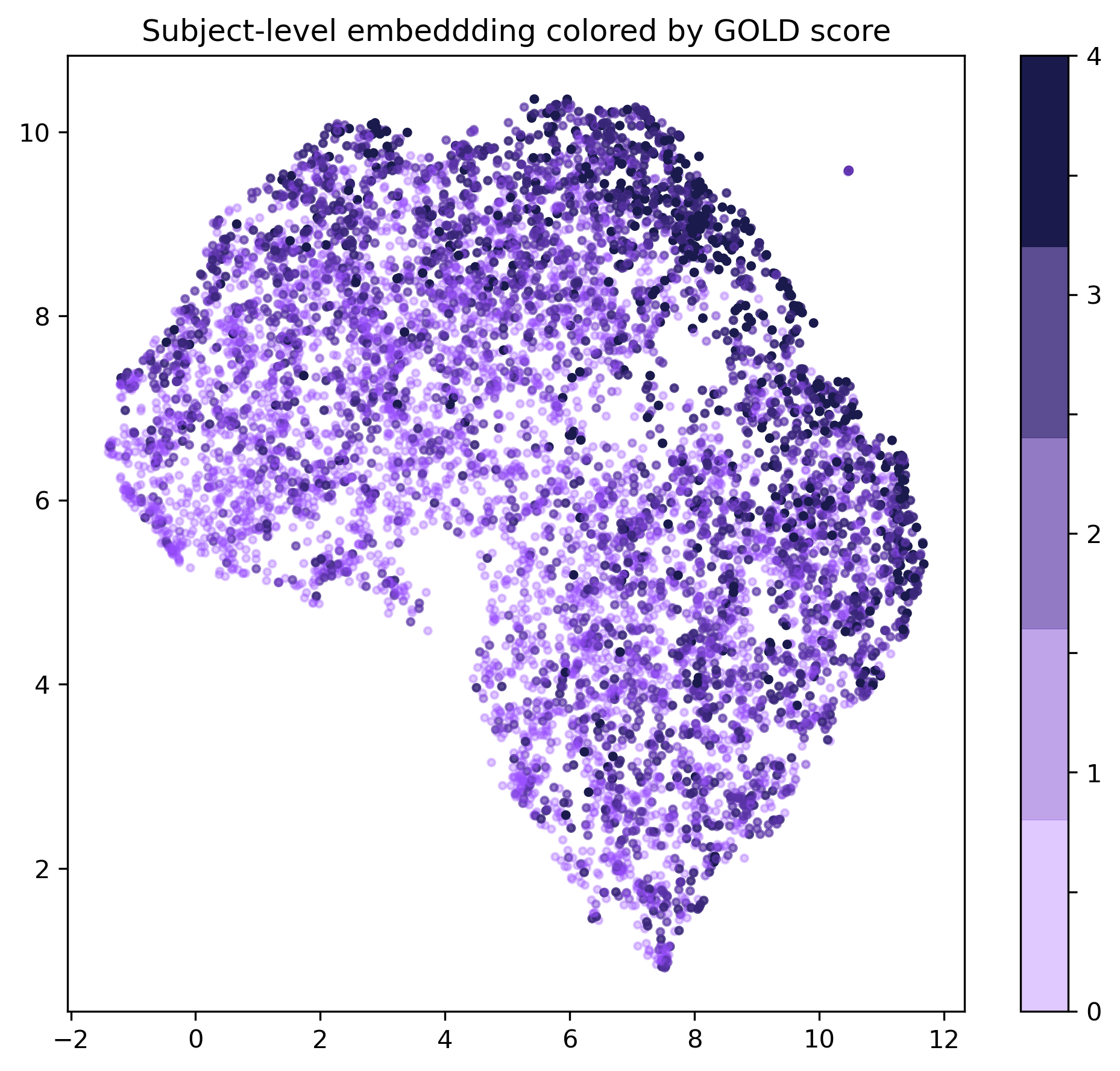}
    \caption{Embedding of subjects in 2D using UMAP. Each dot represents one subject colored by the GOLD score. We can find a trend, from lower-left to upper-right, along which we can see increasing GOLD score. 
    }
    \label{fig:gold}
\end{figure}

To visualize the learned embedding and understand the model's behavior, we use two methods to visualize the model. The first one is embedding visualization,
we use UMAP~\cite{mcinnes2018umap} to visualize the patient-level features extracted on the COPDGene dataset in two dimension.
In Fig~\ref{fig:gold}, we found that subjects with GOLD score of (0,1) and (3,4) are separable under two dimension. But subjects with GOLD score 2 are scattered. It requires further investigation to understanding of embedding pattern of subjects subjects with GOLD score 2. 
In Fig~\ref{fig:gold}, we can find a trend, from lower-left to upper-right, along which we can see increasing GOLD score.

\begin{figure}[t]
\centering
    \includegraphics[width = .49\textwidth]
    {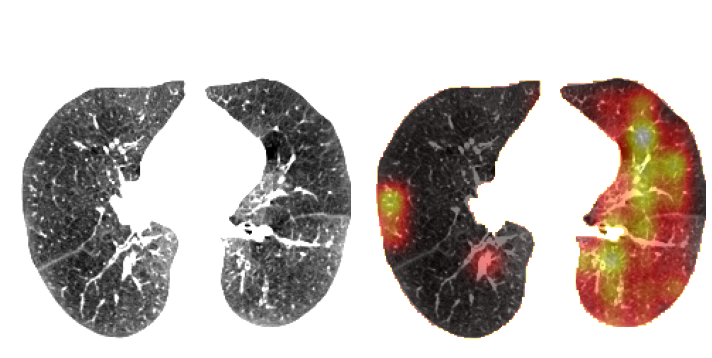}
    \caption{An axial view of the activation map on a $GOLD$ 4 subject in COPDGene dataset. Brighter color indicates higher relevance to the disease severity. The figure illustrates that high activation region overlaps with dark area on  the  right  lung,  where  lung  tissue  is  damaged.
    }
    \label{fig:copd}
\end{figure}

\begin{figure}[t]
\centering
    \includegraphics[width = .49\textwidth]
    {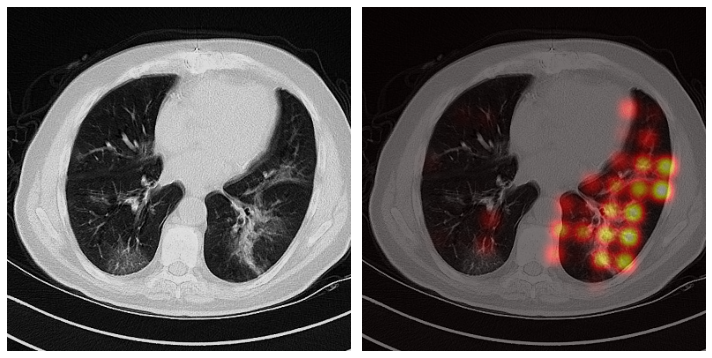}
    \caption{An axial view of the activation heatmap on a COVID-19 positive subject in the COVID-19 CT dataset. Brighter color indicates higher relevance to the disease severity. The figure illustrates that high activation region overlaps with the ground glass opacities.
    }
    \label{fig:covid}
\end{figure}

We use the model explanation method described before to visualize discriminative image regions used by our model for prediction in downstream task. In Fig.~\ref{fig:copd}, we apply the explanation method using the target logit of GOLD score = 4 on a GOLD 4 subject in COPDGene dataset. The dark area on the right lung, where lung tissue is severely damaged, received highest activation value. Figure~\ref{fig:covid} (left) shows the axial view of the CT image of a COVID-19 positive patient, and Figure~\ref{fig:covid} (right) shows the corresponding activation map. The anatomical regions received high activation scores overlap with the peripheral ground glass opacities on the CT image, which is a known indicator of COVID-19. This result suggests that our model can highlight the regions that are clinically relevant to the prediction.

\bibliography{egbib.bib}